\documentclass[10pt,twocolumn,a4paper]{article}
\usepackage[pdftex]{graphics}
\usepackage{graphicx}
\usepackage {layout}
\usepackage{amsmath}
\usepackage{subfigure}
\title{Identification of the ECR zone in the SWISSCASE ECR ion source}
\author{M. Bodendorfer$^{1,2}$, K. Altwegg$^{1}$, P. Wurz$^{1}$, H. Shea$^{2}$ \\  \small{\textbf{1:} Space Research and Planetary Sciences,}  \\ \small{University of Bern, 3012 Bern, Switzerland}  \\ \small{\textbf{2:} Microsystems for Space Technologies Laboratory,} \\ \small{EPFL - Ecole Polytechnique F\'ed\'erale de Lausanne, 1015 Lausanne, Switzerland;}}
\begin{document}

\maketitle

\begin{abstract}
\textbf{The magnetic field of the permanent magnet electron cyclotron resonance (ECR) ion source SWISSCASE located at the \textit{University of Bern} has been numerically simulated and experimentally investigated. For the first time the magnetized volume qualified for electron cyclotron resonance at 10.88 \textit{GHz} and 388.6 \textit{mT} has been analyzed in highly detailed 3D simulations with unprecedented resolution. The observed pattern of carbon coatings on the source correlates strongly with the electron and ion distribution in the ECR plasma of SWISSCASE. Under certain plasma conditions the ion distribution is tightly bound to the electron distribution and can considerably simplify the numerical calculations in ECR related applications such as ECR ion engines and ECR ion implanters.}
\end{abstract}

\section{Indroduction}

The reliable and stable operation of an Electron Cyclotron Resonance (ECR) ion source in producing high intensity beams of highly charged ions depends on the successful design of the magnetic confinement. This confinement enables the ECR process and is responsible for the confinement of high energy electrons for the production of highly charged ions. The resonance criteria for the ECR process is given by Eq. (\ref{ECR}) \cite{Chen general,Geller,bodendorfer}. 

\begin{align}
B_{res} = \frac{\omega \: m_e}{e}
\label{ECR}
\end{align}

$B_{res}$ is the magnetic field fulfilling the resonance condition, $\omega$ is the angular frequency of the applied electric field, $m_e$ is the electron mass and $e$ is the charge of the electron. There is very little that can be done to change the magnetic field geometry after production of the permanent magnets. It is therefore very important to perform reliable and precise numerical simulations of the magnetic field distribution before the production process to guarantee a successful ECR function. 

\begin{figure}[h!]
\unitlength1.0cm
\includegraphics[angle=0, width=0.5\textwidth]{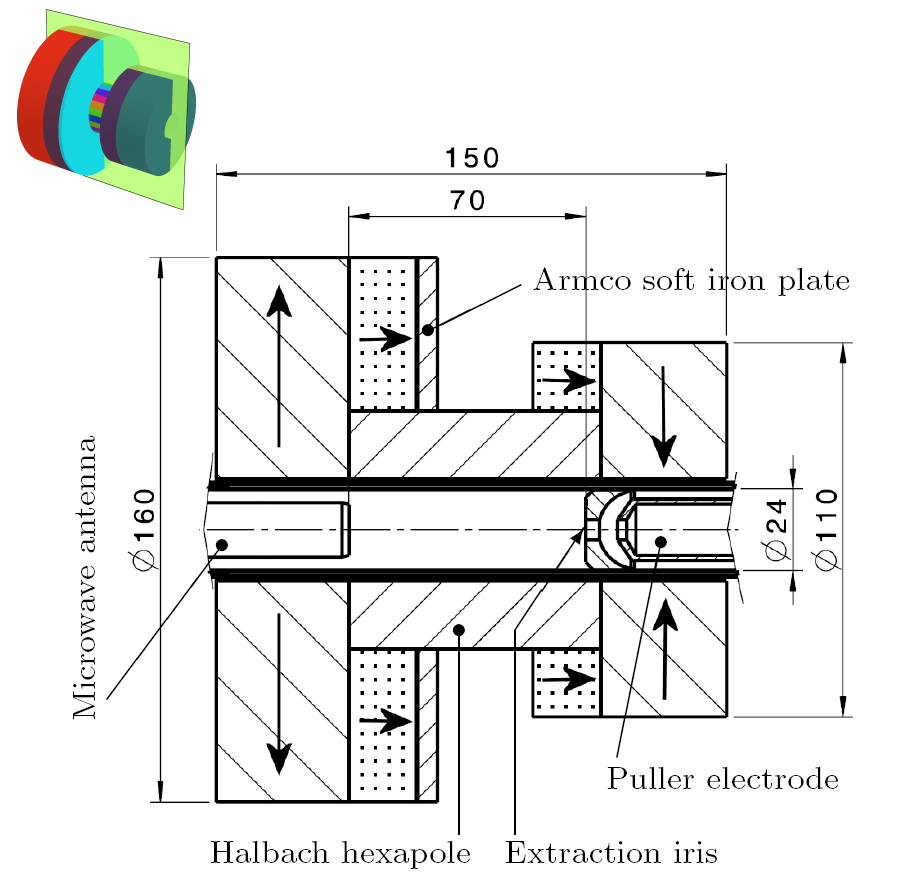}
%\put(-8.0,5.5){\includegraphics[angle=0, width=0.1\textwidth]{3D_fig_cut_YZ.png}}
%\put(-7.5,1){\rotatebox{90}{\footnotesize{Microwave antenna}}}
%\put(-7.2,1.2){\line(3,4){1.0}}
%\put(-6.19,2.55){\circle*{0.1}}
%\put(-3.45,4.8){\rotatebox{0}{\footnotesize{Armco soft iron plate}}}
%\put(-3.55,4.85){\line(-2,-1){0.8}}
%\put(-4.37,4.45){\circle*{0.1}}
%\put(-2.8,0.4){\rotatebox{0}{\footnotesize{Puller electrode}}}
%\put(-2.7,0.68){\line(1,5){0.38}}
%\put(-2.32,2.58){\circle*{0.1}}
%\put(-3.7,-0.25){\rotatebox{0}{\footnotesize{Extraction iris}}}
%\put(-3.6,0.0){\line(0,1){1.8}}
%\put(-3.6,1.8){\vector(2,3){0.6}}
%\put(-6.3,-0.25){\rotatebox{0}{\footnotesize{Halbach hexapole}}}
%\put(-4.1,0.0){\line(0,1){1.8}}
%\put(-4.1,1.8){\circle*{0.1}}
\caption{\textit{A cross section of the relevant components of the operational ECR ion source SWISSCASE at the University of Bern. Arrows indicate the direction of magnetization in the ring shaped permanent magnets. Dimensions in mm.}}
\label{SWISSCASE_cut}
\end{figure}

\begin{figure}[h!]
\unitlength1.0cm
\includegraphics[angle=0, width=0.48\textwidth]{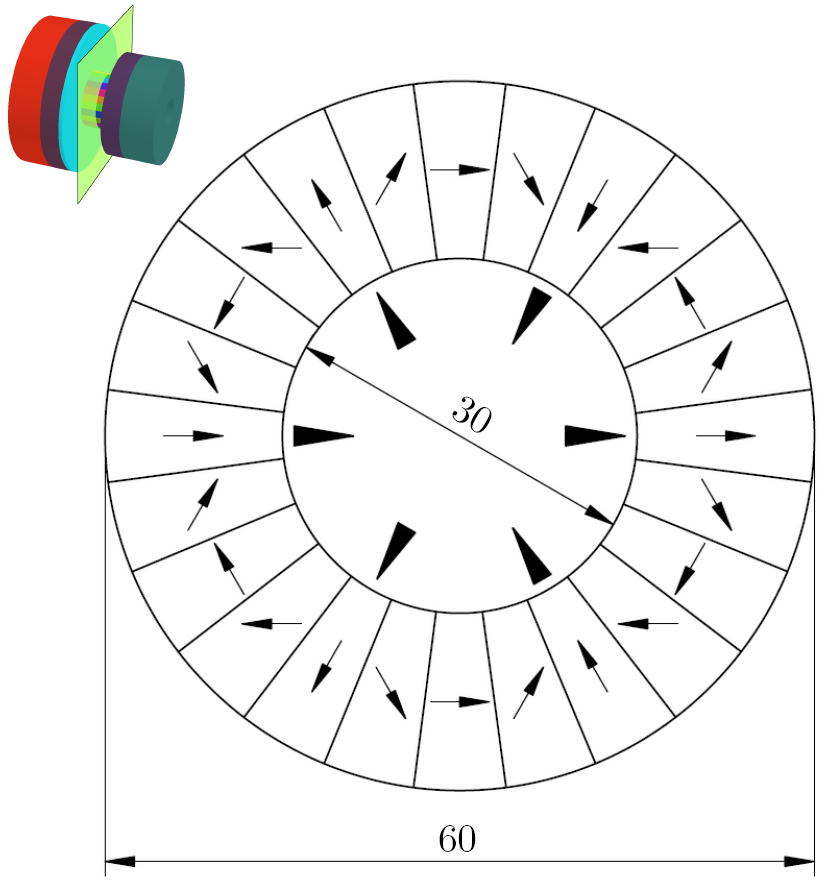}
%\put(-7.8,6.5){\includegraphics[angle=0, width=0.1\textwidth]{3D_fig_cut_XY.png}}
%\put(-4.,0.8){\rotatebox{0}{60}}
%\put(-3.9,4.65){\rotatebox{-30}{30}}
\caption{\textit{A section cut through the Halbach hexapole magnet of SWISSCASE reveals the characteristic alternating orientation of 24 permanent magnet sectors. Bold arrows indicate the resulting main poles inside the hexapole. Dimensions in mm.}}
\label{Halbach}
\end{figure}

In this paper we present the simulation of the newly developed and operational \textbf{SWISSCASE} (\textbf{S}olar \textbf{W}ind \textbf{I}on \textbf{S}ource \textbf{S}imulator for the \textbf{CA}libration of \textbf{S}pace \textbf{E}xperiments) ECR ion source (Figure \ref{SWISSCASE_cut} and \ref{Halbach}) and demonstrate the simulation precision that is necessary for the reliable localization of the ECR process. In SWISSCASE the ring shaped magnets establish the axial confinement featuring rotational symmetry with respect to the beam axis. The Halbach hexapole \cite{halbach} (Figure \ref{Halbach}) establishes the radial electron confinement. 

The combination of the ring shaped magnets and the hexapole requires a full 3D simulation, rather than two 2D simulations, which would require far less computation power. Therefore, a three dimensional simulation has been performed covering the full spatial extent of the magnetic confinement.

%Comparing the magnetic field of SWISSCASE with the simulation of the magnetic field \cite{bodendorfer} in the MEFISTO ECR ion source \cite{AdrianMarti, Hohl, Wurz, Hohl2} reveals differences in structure and smoothness of the ECR qualified plasmoid surface \cite{Bostick}.

In addition to numerically determining the magnetic field distribution, the simulation also allows calculating electron trajectories. In this paper we show that the resulting electron current density and electron charge density coincide with experimentally observed surface coating and sputtering shapes. Further calculations show the ions are bound to the same location as the high energy electron population. This explains the coincidence of the simulated electron distribution and the observed ion surface coating patterns.

In Section \ref{Plasma parameters} we present the calculation of critical plasma parameters to justify the results of the numerical simulation. In Section \ref{Magnetic field} the results of the magnetic field simulation are presented. Section \ref{electron distribution} details the simulated electron distribution followed by the experimentally observed ion coating patterns described in Section \ref{observation}. In Section \ref{relation} the simulation results are compared to our observations.

\section{Plasma Parameters}
\label{Plasma parameters}

Knowledge of the plasma parameters is required for the calculation of mean free path lengths for high energy electrons as well as low energy ions inside the ECR plasma. It is assumed the plasma is dominated by a cold electron population \cite{Hohl}. It is further assumed the plasma frequency to be the cut off frequency for the incident microwaves inhibiting further microwave plasma heating and limiting the cold plasma electron density \cite{Chen general,Geller}. This leads to a stable equilibrium where the cold plasma density is bound to the microwave frequency. Hence the cold plasma density can be estimated by equating the microwave frequency and the plasma frequency: Eq. (\ref{opaque}) \cite{Geller}. All numerical values are for SWISSCASE.

\begin{align}
f_{pe} = f_{MW} = 10.88 \ GHz
\label{opaque}
\end{align}

$f_{MW}$ is the incident microwave frequency and $f_{pe}$ is the plasma frequency. From the plasma frequency one gets the density as \cite{Chen general, Geller}:

\begin{align}
n_{e} = \frac{\omega_{pe}^2 \epsilon_0 m_e}{e^2} = 1.46 * 10^{18} ~ \frac{1}{m^3}
\label{density}
\end{align}

$\omega_{pe}$ is the angular plasma frequency, $\epsilon_0$ the vacuum permittivity, $m_e$ the electron mass and $e$ its charge. 

Because of the acceleration of electrons in the ECR zone, cold electrons are heated to substantial energies, forming the hot electron population, which is responsible for the ionization of atoms and ions.

Lacking our own Bremsstrahlung measurements, which allows the determination of the plasma electron temperature, we assume a temperature of the hot electron population \cite{Geller} of $25\ keV$, which has been measured by R. Friedlin et al. \cite{friedlein} who obtained their measurements by operating an ECR discharge at $7.25\ GHz$. With a value of  $25\ keV$ we obtain the mean free path lengths of the hot electron population according to Table \ref{lambda}. Assuming the same hot electron temperature for SWISSCASE, operated at $10.88\ GHz$, as for the 7.25 GHz source provides a conservative estimate for the hot electron energy and hence for the collision cross section.

%\begin{math}
\begin{table}[h!]
\centering
\begin{tabular}{l|l|l}
Parameter 	& Value & Term \\
\hline
%&&\\
$\sigma_{Spitzer}	$	& $2.05 * 10^{-24}\ m^2$ & $(\frac{e^2}{\epsilon_0 E_{kin}})^2 \frac{1}{2 \pi} ln \Lambda $  \\
%&&\\
\hline
$\lambda_{Spitzer} $ & $3.33 * 10^5\ m $ & $ 1 / n_{e} \sigma_{Spitzer}$ \\
\end{tabular}
\caption{\textit{$\sigma_{Spitzer}$: collision cross section for Spitzer collisions, $ln\Lambda$: Coulomb logarithm, $n_{e}$: plasma electron density, $E_{kin}$: kinetic energy of the electrons.}}
\label{lambda}
\end{table}
%\end{math}

Different formulas for the calculation of mean free path lengths according to different collision mechanisms can be found in the literature (Table \ref{lambda_summary}).

\begin{table}[h!]
\centering
\begin{tabular}{l|l|l}
Coll. mech. 	& $\lambda$ & Ref. \\
\hline
Spitzer 						& $3.33 * 10^5\ m $ 	& 	Geller \cite{Geller} 	\\
\hline
Single $90 ^{\circ}$ & $5.90 * 10^6\ m$ 	& 	Chen \cite{Chen general} 	\\
\hline
Semi empirical			& $3.54 * 10^6\ m$   &   Huba \cite{naval} 	\\
\end{tabular}
\caption{\textit{Mean free path lengths for different collision mechanisms.}}
\label{lambda_summary}
\end{table}

For the source operation with $CO_2$ gas as will be discussed in next sections, the following calculation shows the collision mechanism between electrons and neutrals lead to far shorter mean free path lengths than predicted by electron ion collisions. The neutral population consists of $CO_2$, $CO$, $O_2$ and $O_3$ molecules and of $O$ and $C$ atoms. We simplify the neutral population to a combined total density of $2*n_{CO2}$. It is further assumed this hypothetical population to feature an effective cross section given by a disc with a radius equal to half the length of a $CO_2$ molecule ($116.3\ pm$). This results in an effective cross section of $\sigma_{CO2}' = 4.25 * 10^{-20}\ m^2$. The mean free path can be calculated by Eq. (\ref{CO2_lambda}):

\begin{align}
\lambda_{CO_2} = \frac{1}{2\,n_{CO2}\,  \sigma_{CO2}'}
\label{CO2_lambda}
\end{align}

$\lambda_{CO_2}$ is the mean free path in the $CO_2$ neutral gas background and $n_{CO2}$ its density. Based on fluid and thermodynamical calculations and assuming the neutral gas temperature to be $300\ K$, the neutral gas density in the plasma chamber is $4.805 * 10^{18}\ 1/m^3$ resulting in $\lambda_{CO_2} = 4.90\ m$. Hence the limiting mechanism for the mean free path is given by electron collisions with neutrals rather than electron collisions with electrons or ions. 

Another effect that must be considered is radial electron loss across the magnetic field lines by collision induced diffusion. A simplified expression for the diffusion process of the hot electrons is given by Eq. \ref{diff} \cite{Chen general}.

\begin{align}
\Gamma_e = -\mu \ n_e \ E - D_e \ \nabla n_e
\label{diff}
\end{align}

$\Gamma_e$ is the loss rate of the electrons per time and area unit. $\mu$ is the electron mobility, $n_e$ the electron density, $E$ the electric field and  $D_e$ the electron diffusion coefficient. $D_e$ is given by Eq. \ref{De} \cite{Geller}.

\begin{align}
D_e = r_L^2 \ \nu_{e-n}
\label{De}
\end{align}

$r_L$ is the Larmor radius of the hot electrons and $\nu_{n-e}$ the dominating collision frequency between electrons and neutrals as shown in the previous paragraph. With $r_L = 1.37$ $mm$ and $\nu_{e-n} = 1.92*10^{7}$ $1/s$ we obtain $D_e = 36.1$ $m^2/s$.

Furthermore $\mu$ is related to $D_e$ by the Einstein-Smoluchowski equation (Eq. \ref{einstein}) \cite{Chen general}.

\begin{align}
\mu = \frac{e \ D_e}{K \ T}
\label{einstein}
\end{align}

$K$ the Boltzman constant and $T$ the hot electron temperature. With $T=25$ $keV$ we obtain $\mu = 1.44*10^{-3}$ $m^2/sV$. We can now compare the two terms on the right side of Eq. \ref{diff}:

\begin{align}
\mu \ n_e \ E = 5.28 * 10^{17} \ \frac{1}{sm^2}
\label{comp1}
\end{align}

\begin{align}
D_e \ \nabla n_e = 4.39 * 10^{21} \ \frac{1}{sm^2}
\label{comp2}
\end{align}

$E$ is the electric field assumed to be the plasma potential of 3 $V$ divided by the plasma chamber radius $R=12$ $mm$. $\nabla n_e$ has been estimated by the electron density $n_e$ divided by the plasma chamber radius. From the obtained values we can see the diffusion process is dominated by the term $D_e \ \nabla n_e$ rather than the term $\mu \ n_e \ E$. This allows simplifying the calculation of the number of collisions with neutrals an electron undergoes before being lost by diffusion (Eq. \ref{ncoll}).

\begin{align}
n_{coll} = T_D \ \nu_{e-n}
\label{ncoll}
\end{align}

$T_D$ is the mean time the electrons spend in the confinement before being lost by the discussed diffusion process. $\nu_{e-n}$ is the electron collision frequency with neutrals. We can write Eq. \ref{ncoll} as Eq. \ref{ncoll1}:

\begin{align}
n_{coll} = \frac{n_e \ V_{Plasma} \ \nu_{e-n}}{\dot{n}_{De} \ A_{Plasma}}
\label{ncoll1}
\end{align}

$V_{Plasma}$ is the plasma volume, $\dot{n}_{De}$ the electron loss rate equal to the right hand side of Eq. \ref{comp2} and $A_{Plasma}$ the plasma surface. We approximate the geometry of the ECR plasma by a cylinder with length and diameter of the ECR plasmoid. Further introducing $\epsilon_{V\!\!A} = V_{Plasma} / A_{Plasma}$ as the ratio of the plasma volume and the plasma surface leads to Eq. \ref{ncoll2}:

\begin{align}
n_{coll} = \frac{n_e \ \epsilon_{V\!\!A} \ \nu_{e-n} \ R}{D_e \ n_e} = \frac{\epsilon_{V\!\!A} \ R }{r_L^2}
\label{ncoll2}
\end{align}

With our values one obtains $n_{coll} = 19.1$. This means hot electrons with a kinetic energy of 25 $keV$ undergo about 19 collisions with neutrals before being expelled from the plasma by diffusion across the magnetic field lines. Relating to the CO$_2$ gas operation discussed further down a similar calculation for ions yields an average of 22 collisions with neutrals before a singly charged carbon ion is lost by diffusion.

We conclude from this section that the calculated mean free paths of the hot electrons are long enough to fit many times into the plasma container and that diffusion across the magnetic field lines is not impeding the trajectories of the hot electrons because many collisions are needed to complete a successful diffusion loss mechanism. Hence the trajectory integration based on the simulated magnetic field presented in the next section is valid. Furthermore, considering diffusion across magnetic field lines, ions are approximately as well confined as electrons in terms of number of collisions before loss.

\section{The magnetic field}
\label{Magnetic field}

For all magnetic simulations of SWISSCASE, the software TOSCA by \textit{vectorfields} \cite{bodendorfer}, a finite element solver, is used. A summary of the essential parameters of the finite element model are presented in Table (\ref{FEM_summary}).

\begin{table}[h!]
\centering
\begin{tabular}{l|r}
number of linear elements & $4.7*10^6$ \\
\hline
number of quadratic elements & $4.8*10^5$\\
\hline
number of nodes & $1.6*10^6$ \\
\hline
number of edges & $6.1*10^6$ \\
\hline
RMS & $1.033$  \\
\hline
distortion of worst element & $1.40*10^{-3}$
\end{tabular}
\caption{\textit{Some characteristics of the finite element model used for the numerical calculation of the magnetic field.}}
\label{FEM_summary}
\end{table}

\begin{figure}[h!]
\center
\unitlength1.0cm
\includegraphics[angle=0, width=0.4\textwidth]{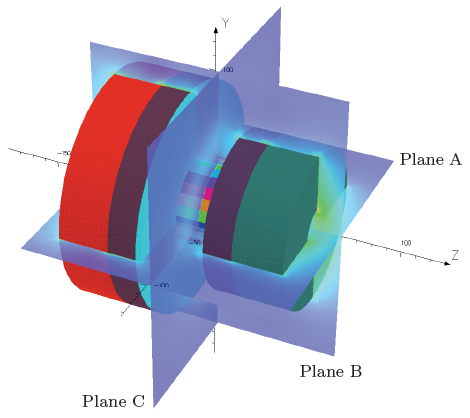}
%\put(-6.4,0.1){\rotatebox{0}{\footnotesize{Plane C}}}
%\put(-1.15,4.1){\rotatebox{0}{\footnotesize{Plane A}}}
%\put(-2.8,0.6){\rotatebox{0}{\footnotesize{Plane B}}}
\caption{\textit{All magnetic parts of the finite element model have been meshed carefully according to size and relevance thereby optimizing the simulation efficiency.}}
\label{FEM}
\end{figure}

Figure \ref{FEM} shows the finite element model used for the magnetic field simulation. Cut planes have been introduced for the visualization of section views.

The axial magnetic field has been measured using a Hall probe. The simulation result did not deviate from the measurement more than $2.8\ \%$ at any point along the $z-axis$ (see Figure \ref{FEM}). The maximal error near the center of the electron cyclotron resonance zone was $0.97\ \%$.
Figures \ref{Cut_A}, Figures \ref{Cut_B} and \ref{Cut_C} present section views of the magnetic model defined by the cut planes introduced in Figure \ref{FEM}.

\newcommand{\bildbreite}{0.4}

\begin{figure}[h!]
 \begin{center}
   \subfigure[\textit{}]
             {
              \includegraphics[angle=0, width=\bildbreite \textwidth]{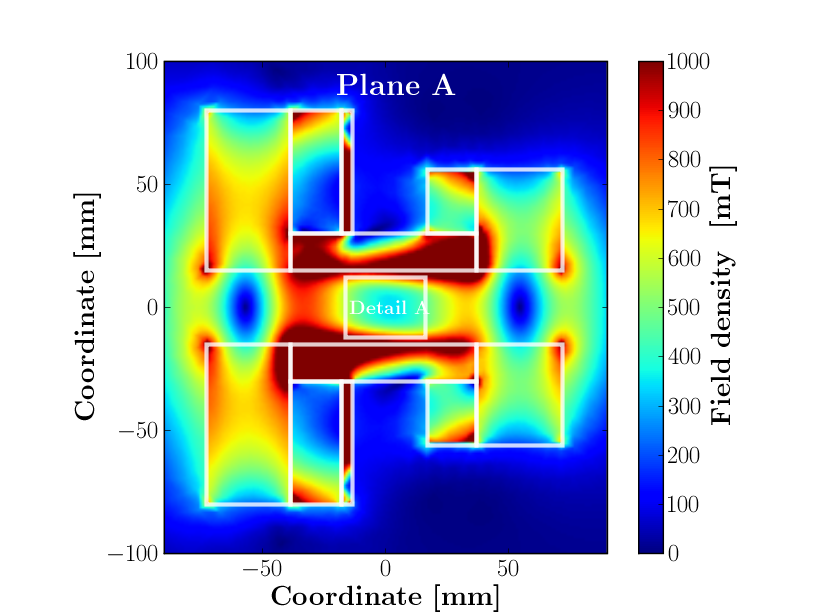}
             }
  % \vspace{-40pt}
   \subfigure[\textit{}]
             {
              \includegraphics[angle=0, width=\bildbreite \textwidth]{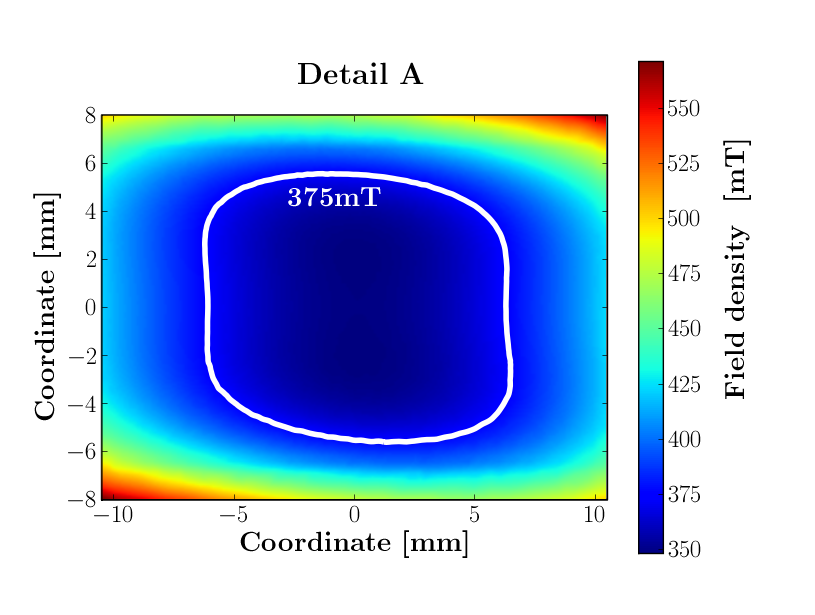}
             }
   %\vspace{-20pt}
   \caption{\textit{Simulated magnetic field in plane A. For the purpose of orientation the outlines of the permanent magnet structure and the Armco soft iron plate are represented by white lines in Figure a. Figure (b) features a detailed view of the center section. The ECR iso contour line is highlighted, revealing the topography of the ECR qualified surface.}}
   \vspace{-20pt}
    \label{Cut_A}
  \end{center}
\end{figure}

\begin{figure}[h!]
 \begin{center}
   \subfigure[\textit{}]
             {
              \includegraphics[angle=0, width=\bildbreite \textwidth]{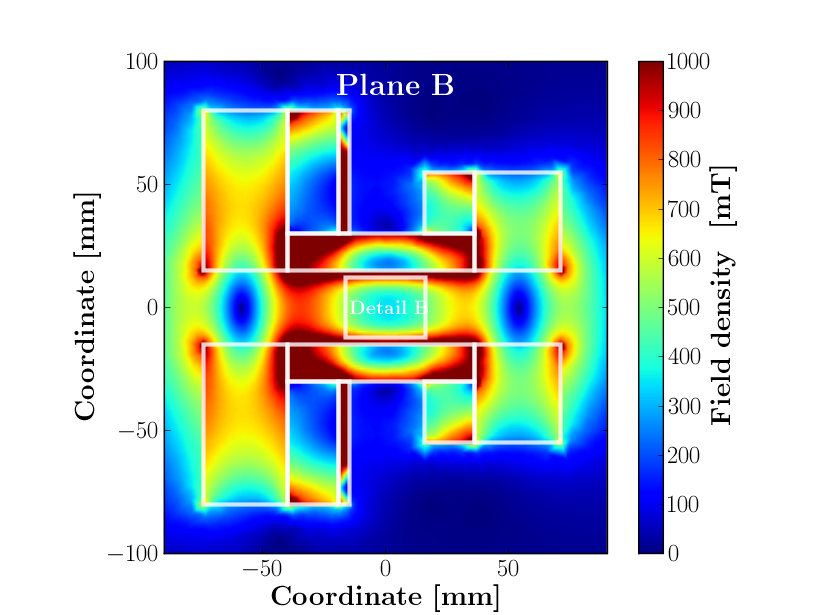}
             }
   \subfigure[\textit{}]
             {
              \includegraphics[angle=0, width=\bildbreite \textwidth]{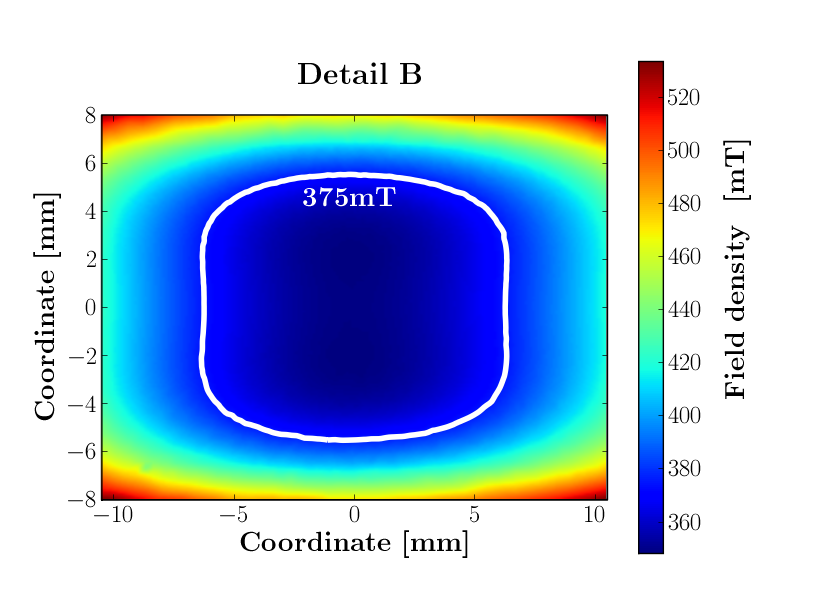}
             }
   \caption{\textit{Simulated magnetic field in plane B.}}
   \vspace{-30pt}
    \label{Cut_B}
  \end{center}
\end{figure}

\begin{figure}[h!]
 \begin{center}
   \subfigure[\textit{}]
             {
              \includegraphics[angle=0, width=\bildbreite \textwidth]{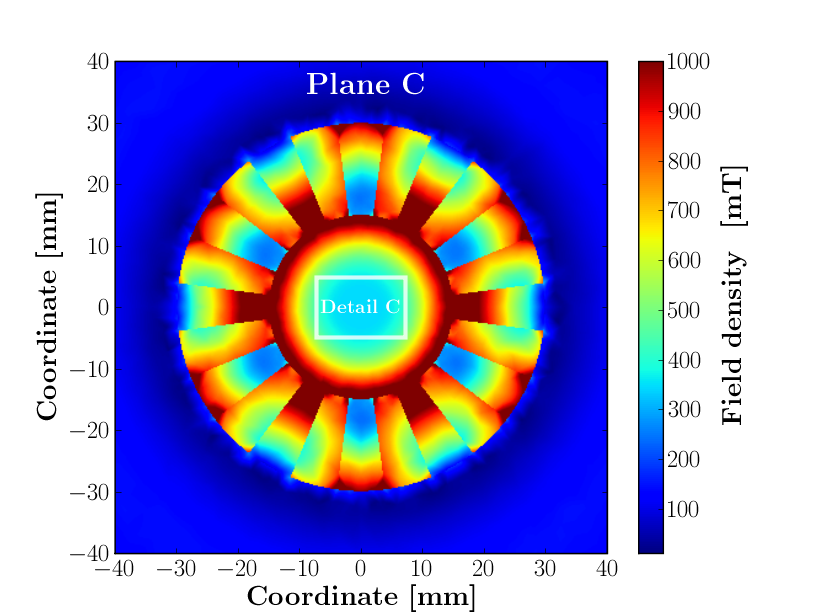}
             }
   \subfigure[\textit{}]
             {
              \includegraphics[angle=0, width=\bildbreite \textwidth]{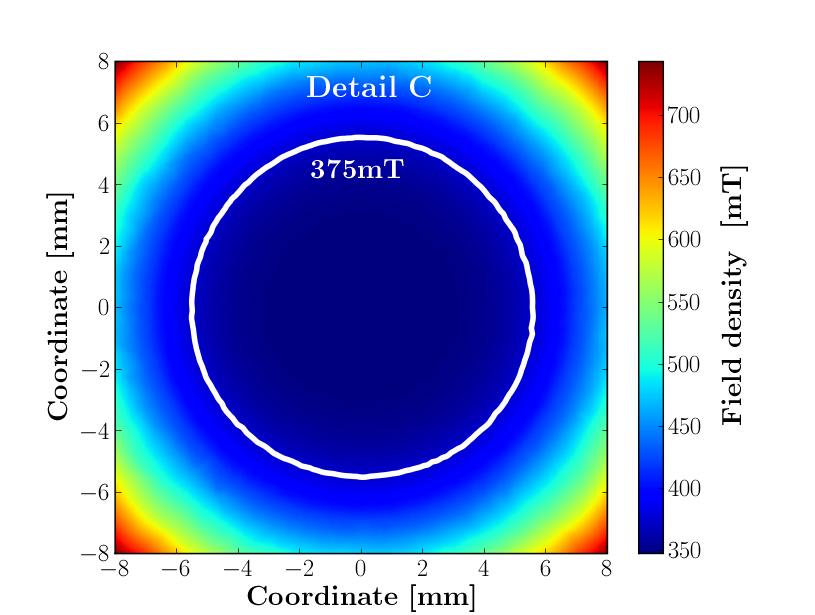}
             }
   \caption{\textit{Simulated magnetic field in plane C. Figure (b) features a detailed view of the center section of Figure (a). The ECR iso contour line appears circular.}}
   \vspace{-20pt}
    \label{Cut_C}
  \end{center}
\end{figure}

The ECR qualified surface, called plasmoid, presents itself compact and point symmetrically deformed as expected from the permanent magnet setup. Figure \ref{Cut_A}b shows, despite its larger size than the permanent magnetic ring to the right (positive z-coordinate), the permanent magnet ring to the left (negative z-coordinate) does not deform the iso contour line in any asymmetric way. However the difference in size of the larger permanent magnet ring shifts the center of the plasmoid along the z axis toward the larger permanent magnetic ring (in Figure \ref{Cut_A}b to the left) without distorting its overall shape on a larger scale.

\section{The electron distribution}
\label{electron distribution}

In addition to the solution of the magnetic model, electron trajectories have been calculated using the finite element model trajectory integrator \cite{bodendorfer} SCALA by \textit{vectorfields}. The initial velocities of the particle trajectories were chosen to be anisotropically distributed. The electron velocity vector components perpendicular to the local magnetic field lines are distributed with a temperature of $25\ keV$. The velocity vector components parallel to the local magnetic field are distributed with a temperature of $2\ eV$ \cite{Hohl}, the temperature of the cold electrons. This anisotropic velocity distribution complies with the ECR heating mechanism with the same anisotropic orientation which is perpendicular to the local magnetic field lines \cite{Chen general}. In total 36328 trajectories have been launched for a simulation run. The presence of space charge caused by the plasma electrons has been modeled by introducing a charged sphere with the same diameter of the plasmoid and a potential of $-3\ V$ \cite{Golo, Hohl, Bostick}. 

\begin{figure*}
 \begin{center}
   \subfigure[\textit{Plasmoid center}]
             {
              \includegraphics[angle=0, width=0.23\textwidth]{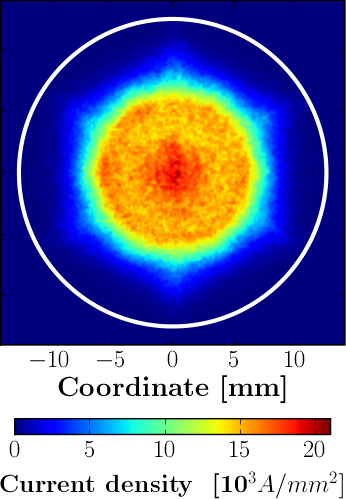}
              %\put(-104,-3){\includegraphics[angle=0, width=0.23\textwidth]{JBeam_scale.png}}
             }
   \subfigure[\textit{$5\ mm$ from the center}]
             {
              \includegraphics[angle=0, width=0.23\textwidth]{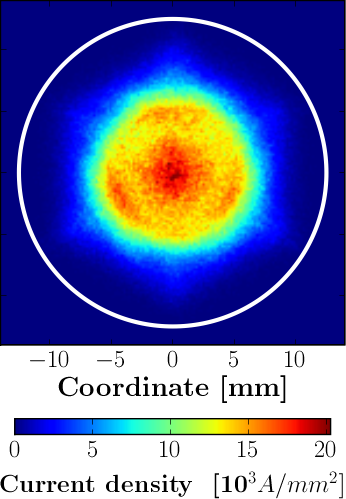}
              %\put(-104,-3){\includegraphics[angle=0, width=0.23\textwidth]{JBeam_scale.png}}
             }
   \subfigure[\textit{$10\ mm$ from the center}]
             {
              \includegraphics[angle=0, width=0.23\textwidth]{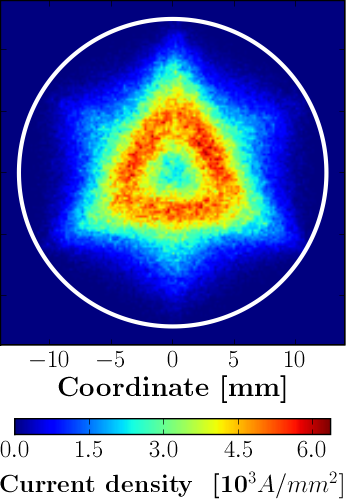}
              %\put(-104,-3){\includegraphics[angle=0, width=0.23\textwidth]{JBeam_scale.png}}
             }
   \subfigure[\textit{$15\ mm$ from the center.}]
             {
              \includegraphics[angle=0, width=0.23\textwidth]{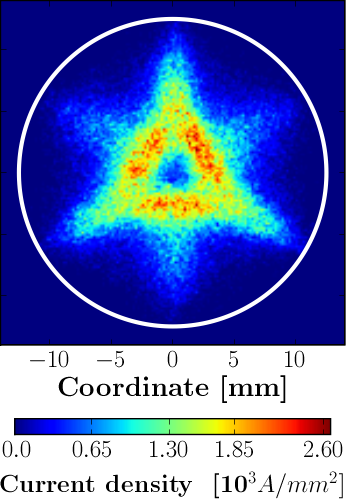}
              %\put(-104,-3){\includegraphics[angle=0, width=0.23\textwidth]{JBeam_scale.png}}
             }
   \caption{\textit{Electron current density maps at different positions along the optical axis. Note the increasing similarity with surface coating triangles (see Figure \ref{triangle}) toward the antenna tip at $15\ mm$ offset from the plasmoid center. The white ring represents the plasma chamber wall (diameter 24 mm). Dimensions in $mm$.}}
   \vspace{-20pt}
    \label{J_series}
  \end{center}
\end{figure*}

The magnetic field distribution of the MEFISTO ECR ion source located at the University of Bern has been simulated by Bodendorfer at al. \cite{bodendorfer} and matches the observed surface coating triangles in MEFISTO. The simulated magnetic field distribution of SWISSCASE does not resemble the respective surface coating triangles in SWISSCASE to the same extent. However, a closer look at the electron distribution instead of the magnetic field distribution, suggests a strong link between electron distribution and surface coatings in SWISSCASE. Figures \ref{J_series}a-d illustrate how the shape of the simulated electron current density distribution changes from a hexagon at the plasmoid center into a triangular shape at the tip of the microwave antenna, where a triangular-shaped surface coating has been observed (Figure \ref{triangle}) after the operation of the SWISSCASE ECR ion source \cite{bodendorfer}. This transformation of the electron density maps represents the changing cross section of the ECR zone while the cut plane is moved from the plasmoid center toward the antenna. Note due to the geometry of the magnetic field, the plasmoid and the electron density distribution are point symmetric with respect to the plasmoid center.

In the next section we present the observed coating pattern and a spectral analysis of it which shows the coating consists indeed of particles found in the plasma.

\section{The observed surface coating pattern}
\label{observation}

\begin{figure}[h!]
\includegraphics[angle=0, width=0.5\textwidth]{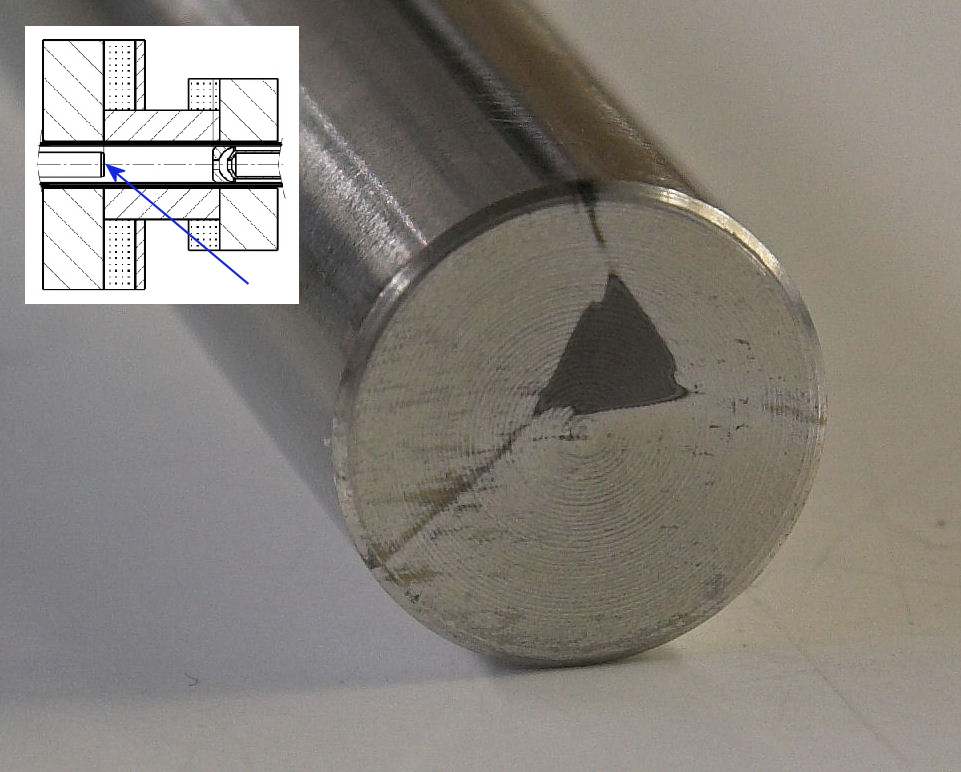}
\caption{\textit{Photograph of the microwave antenna of SWISSCASE after operation ($CO_2$, 100 W input power, 10.48 GHz), published by Bodendorfer at al. in \cite{bodendorfer}. A dark triangular patch consisting mostly of carbon is clearly visible, which was not present before the operation. The shape of the deposited coating matches the electron density map, as explained in the text and in Figure \ref{J_series}. The pattern is displaced from the optical axis due to a misalignment of the microwave antenna caused by the antenna's length and weight. The misalignment did not cause any measurable decrease in the ion beam performance. The antenna has a diameter of 12 mm and the side length of the triangle measures 4.2 mm.}}
\vspace{-20pt}
\label{triangle}
\end{figure}

Figure \ref{triangle} shows a typical surface coating triangle resulting from the exposure directly or indirectly to the ECR plasma. In this particular case the observed surface coating triangle was at the tip of the microwave antenna (see Figure \ref{SWISSCASE_cut}) in the SWISSCASE facility after the operation of a $CO_{2}$ plasma. In some cases the observed pattern is a sputtered area rather than a surface coating. This may originate from the difference in particle energies the surface is exposed to. The observed surface coatings have been exposed directly to the plasma without any kind of extraction or post acceleration of the plasma particles. In this case the particle energies may have been low enough to enable a deposition process of the neutralized carbon atoms to form the observed surface coating. On the other hand sputtering shapes are observed after the extraction and post acceleration where the involved particles have energies many orders of magnitude larger than they had inside the plasma. In this case the particle energy might be too large for a continuous deposition process and hence sputtering occurs.

The antenna tip was chemically analyzed to determine the composition of the surface coating. The result is summarized in Table \ref{spektral_analysis}.

\begin{table}[h!]
\centering
\begin{tabular}{l|l|l}
Element & Weight [\%] & Abundance [\%] \\
\hline
Carbon & $90.61$ & $93.04$\\
\hline
Oxygen & $7.36$ & $5.67$ \\
\hline
Fluorine & $1.41$ & $0.91$ \\
\hline
Neon & $0.62$ & $0.38$  \\
\hline
Total & $100$ & $100$
\end{tabular}
\caption{\textit{Composition of the surface coating in mass and abundance fraction.}}
\label{spektral_analysis}
\end{table}

The dominance of carbon in the coated surface suggests a deposition process originating from the plasma because the plasma mainly consists of carbon and oxygen neutrals and ions. Oxygen is found as a minor fraction only because it mainly recombines with other oxygen and carbon radicals and molecules to form molecular oxygen, ozone, carbon monoxide or carbon dioxide. All of these recombination products are gaseous and therefore not likely candidates for a deposition process. FC40, a liquid composed of Fluorine and Carbon, N-(C$_3$F$_7$)$_3$, is used as cooling fluid enveloping the plasma chamber. Fluorine combines well with metals. Therefore, the Fluorine traces in the spectral analysis may originate from cooling fluid diffusing in small amounts through the copper plasma chamber wall. Neon found in the spectral analysis is a remainder of previous Neon plasma operations in the same plasma chamber.

\section{Relation between simulation and observation}
\label{relation}

In this section we explain the relation between the observed surface coating pattern and the simulated electron distribution. 

As discussed in Section \ref{Plasma parameters} the electron mean free path is limited to $4.9\ m$ by collisions with neutrals. The mean free path is therefore far longer than the ECR plasmoid. This in turn potentially allows ionization processes anywhere inside the plasma container. However the magnetic simulation and the numerical trajectory integration show that the electrons are confined to a limited space inside the confining field (see Figures \ref{J_series}a-d for the electron density profile). 

To investigate the spatial distribution of the ions we make the approximation that the ion-neutral cross section equals to twice the value of the hot electron-ion scattering cross section ($8.5*10^{-20}\ m^2$) and a neutral density equal to the value presented in Section 2 of this article ($4.805*10^{18}$ $1/m^3$). This leads to a mean free path length for ions of $0.8\ m$ which is 62 times the length of the ECR plasmoid. This suggests the ions would completely fill all of the vessel volume resulting in an uniform density distribution. However in a magnetic field of $388.6\ mT$ singly charged carbon ions with a kinetic energy of $2\ eV$ gyrate with a Larmor radius of $1.82\ mm$ around the magnetic field lines on which they were created. The trajectories of the carbon ions are therefore restricted by the magnetic confinement field rather than the limited geometry of the plasma vessel. Hence the ions do indeed travel for a long mean free path length relative to the ECR plasmoid dimensions. But the ions are prevented by the magnetic field from filling all the vessel volume and are confined to a restricted space similar to the electron distribution presented in the last section. 

This explains the good agreement between the observed surface coating triangle and the simulated electron distribution because the carbon ions do not travel far from the region of ionization defined by the presence of high energy electrons.

\section{Conclusion}
\label{Conclusion}

We simulated in detail the 3D magnetic field distribution of the SWISSCASE ECR ion source. The simulation of high energy electron trajectories based on the same magnetic model leads to shapes nearly identical to the observed ion surface coating triangle. This implies the ion distribution is related to the high energy electron distribution. This is justified by the magnetic field parameters, the ion mean free path lengths and similar Larmor radii and diffusion parameters of hot electrons and singly charged carbon ions in SWISSCASE.

Furthermore we generalize that lightweight ions with Larmor radii and mean free path lengths similar to those of a hot electron population can be considered as spatially bound to the latter. This can significantly simplify the calculation of the ion distribution in ECR related applications such as ECR ion engines and ECR ion implanters.

\section{Acknowledgments}
\label{Acknowledgments}

We thank Muhamed Niklaus for the spectral analysis of the antenna tip sample. We gratefully acknowledge the financial support of this project by the Swiss National Science Foundation, grant number 200020-116189.


\begin{thebibliography}{99}
\bibitem{Chen general} Francis F. Chen, Introduction to Plasma Physics and Controlled Fusion, Volume 1: Plasma Physics, Second Edition, Plenum Press, 1984.
\bibitem{Geller} R. Geller, Electron Cyclotron Ion Sources and ECR Plasmas, Institute of Physics Publishing Bristol and Philadelphia, 1996.
\bibitem{naval} J.D. Huba, NRL PLASMA FORMULARY, Beam Physics Branch, Plasma Physics Devision, Naval Research Laboratory, Washington, DC 20375, 2007.
\bibitem{bodendorfer} M. Bodendorfer, Field structure and electron life times in the MEFISTO electron cyclotron resonance ion source, Nuclear Instruments and Methods in Physics Research B 266, p. 820-828, 2008.
\bibitem{friedlein} R. Friedlein, Experimental study of the hot and warm electron populations in an electron cyclotron resonance argon–oxygen–hydrogen plasma
Cyclotron Resonance Ion Source, Nuclear Instrumentes and Physics research B, Volume 266, Issue 5, p. 820-828, March, 2008.
%\bibitem{Zschornack} V.P. Ovsyannikov, G. Zschornack \emph{Precision adjustment of an ECR plasma chamber relative to the magnetic confinement field using strong focused electron beams, Nuclear Instruments and Methods in Physics Research, A 416 (1998) 18-22}
\bibitem{AdrianMarti} A. Marti, et al., Calibration facility for solar wind plasma instrumentation, Review of Scientific Instruments, Volume 72, Number 2, February, 2001.
%\bibitem{Liehr} M. Liehr, Ph.D. thesis, University of Giessen, Germany, 1992
%\bibitem{Liehr_Trassl} M. Lier, R. Trassl, M. Schlapp, E. Salzborn \emph{A low power 2.45 GHz ECR ion source for multiply charged ions, Review of Scientific Instruments, 63 (1992) April 2541}
\bibitem{Golo} K.S. Golovanivsky and G. Melin, A study of the parallel energy distribution of lost electrons from the central plasma of an electron cyclotron resonance ion source, Review of Scientific Instruments, Volume 63, Number 4, April, 1992.
\bibitem{Hohl} Markus Hohl, Peter Wurz, Peter Bochsler, Investigation of the density and temperature of electrons in a compact 2.45 GHz electron cyclotron resonance ion source plasma by x-ray measurements, PLASMA SOURCES SCIENCE AND TECHNOLOGY, Number 14, p. 692-699, 2005.
\bibitem{Bostick} Winston H. Bostick, Experimental Study of Ionized Matter Projected across a Magnetic Field, Phys. Rev. 104, p. 292-299, 1956.
%\bibitem{shirkov_1} G. D. Shirkov \emph{A classical model of ion confinement and losses in ECR ion sources, PLASMA SOURCES SCIENCE AND TECHNOLOGY, page 250, Number 2, 1993}
%\bibitem{shirkov_2} G. D. Shirkov \emph{Electron and ion confinement conditions in the open magnetic trap of ECR ion sources, CERN-PS-94-13, 11th June 1994}
\bibitem{Wurz} P. Wurz, A. Marti, P. Bochsler, New test facility for solar wind instrumentation, Helvetica Physica Acta 71, 1998.
\bibitem{Hohl2} M. Hohl, Ph.D. thesis, Design, setup, characterization and operation of an improved calibration facility for solar plasma instrumentation, University of Bern, Switzerland, 2002.
\bibitem{web_NIST} National Institute of Standards and Technology, $http://physics.nist.gov/PhysRefData/Ionization/$
\bibitem{halbach} K. Halbach, Design of permanent magnet multipole magnets with oriented rare earth cobalt materials, Nuclear Instruments and Methods 169, p. 1-10, 1980.
\end{thebibliography}
\end{document}